\documentclass[a4paper]{jpconf}
\usepackage{graphicx}
\usepackage{amssymb}
\usepackage{amsmath}
\newcommand{\beq}{\begin{eqnarray}}
\newcommand{\eeq}{\end{eqnarray}}
\newcommand{\open}{\sphericalangle}

\newcommand{\ovl}{\overline}
\newcommand{\nn}{\nonumber}
\usepackage{bm}

\begin{document}
\title{Dihadron fragmentation functions and their relevance for transverse spin studies}


\author{A. Courtoy$^{1}$, A. Bacchetta$^{1, 2}$ and M. Radici$^{1}$}

\address{$^{1}$ INFN-Sezione di Pavia,  27100 Pavia, Italy.}
\address{$^{2}$ Dipartimento di Fisica Nucleare e Teorica, Universit\`a di Pavia,   27100 Pavia, Italy.}

\ead{aurore.courtoy@pv.infn.it}


\begin{abstract}
Dihadron fragmentation functions describe the probability that a quark fragments into two hadrons plus other undetected hadrons. In particular, the so-called interference fragmentation functions describe the azimuthal asymmetry of the dihadron distribution when the quark is transversely polarized. They can be used as tools to probe the quark transversity distribution in the nucleon. Recent studies on unpolarized and polarized dihadron fragmentation functions are presented, and we discuss their role in  giving insights into transverse spin distributions.

\end{abstract}

\section{Introduction}

Our knowledge on the hadron structure is  incomplete.  We know that the Parton Distribution Functions (PDFs) describe the one-dimensional structure of hadrons. At leading order, the PDFs are three: number density, helicity and transversity. However the experimental knowledge on the latter is rather poor as it is a chiral-odd quantity not accessible through fully inclusive processes.
Semi-inclusive production of two hadrons~\cite{Collins:1993kq, Jaffe:1997hf} offers an alternative way to access transversity, 
where the chiral-odd partner of transversity is represented by the Dihadron Fragmentation Functions (DiFF) $H_1^{\open}$~\cite{Radici:2001na}, which relates the  transverse spin of the quark to the azimuthal orientation of the two-hadron plane.
Since the transverse momentum of the hard parton is integrated out, the cross section can be studied in the context of collinear factorization. This peculiarity is an advantage over the $p_{T}$-factorization framework, where the cross sections involve  convolutions of the relevant functions instead of simple products.

The transversely polarized DiFF has been computed only in a spectator model~\cite{Bacchetta:2006un}. Recently, the HERMES collaboration has
reported measurements of the asymmetry containing the product
$h_1 H_1^{\open}$~\cite{Airapetian:2008sk}. The COMPASS collaboration has  presented analogous preliminary results~\cite{Martin:2007au}.  
The BELLE collaboration has also  presented preliminary measurements of  the azimuthal asymmetry in $e^+e^-$ annihilation related to the DiFF~\cite{Vossen:2009xz}. 

Our present goal is to extract transversity through this channel. To this end, we need an expression for  the chiral-odd DiFF $H_1^{\open}$ obtained from $e^+e^-$ data. This in its turn requires a knowledge of the unpolarized DiFF $D_1$.
Hence, as a first step,  we present here a parameterization of the unpolarized DiFF  $D_1$ as given from the Monte Carlo generator (MC) of the BELLE collaboration.

\section{Two-hadron Inclusive DIS: towards Transversity}

We consider the SIDIS process 
$e(l)+N^\uparrow(P) \rightarrow e(l')+ \pi^+(P_1)+ \pi^-(P_2)+ X$,  where the 
momentum transfer $q=l-l'$ is space-like, with $l,l'$, the 
lepton momenta before and after the scattering. The two pions coming from the fragmenting quark have  momenta
$P_1$ and $P_2$, respectively, and invariant mass $M_h$, which is considered to be much smaller than the hard scale of the process. We introduce the vectors $P_h=P_1+P_2$ and $R=(P_1-P_2)/2$. We describe a 4-vector $a$ as $[a^-,a^+,a^x,a^y]$,  i.e. in terms of its light-cone 
components $a^\pm = (a^0 \pm a^3)/\sqrt{2}$ and its transverse spatial components. 
We introduce the light-cone fraction $z= P_h^-/k^-$.
 $P$ is the momentum of the nucleon target with mass $M$. We refer to Refs.~\cite{Bacchetta:2006un, Radici:2001na} for details and kinematics.

The spin asymmetry $A_{UT}^{\sin(\phi_R^{} + \phi_S^{})\,\sin\theta}(x,y,z,M_h^2)$ is related to an asymmetric modulation of 
pion pairs in the angles $\phi_S^{}$ and $\phi_R^{}$, which represent the 
azimuthal orientation with respect to the scattering plane of the target 
transverse polarization and of the plane containing the pion pair momenta, 
respectively. The polar angle $\theta$ describes the orientation of $P_1$, in the 
center-of-mass frame of the two pions, with respect to the direction of $P_h$ 
in the lab frame. The asymmetry  is expressed as
\begin{equation} 
A_{UT}^{\sin(\phi_R^{} + \phi_S^{})\,\sin\theta}(x,y, z,M_h^2) \varpropto
-
\frac{|\bm{R}|}{M_h}\,  
\frac{\sum_q e_q^2\,h_1^q(x)\ H_{1,q}^{\open sp}(z,M_h^2)}
     {\sum_q e_q^2\,f_1^q(x)\  D_{1,q}^{ss+pp}(z,M_h^2)} \quad , 
\label{eq:asydis}
\end{equation} 
where the $x$-dependence is given by the PDFs only.  The $z$ and $ M_h$ dependence are governed by the DiFFs whose functional form we need to determine. The  procedure allowing us to give the required parameterizations for the DiFFs is detailed in the following sections.

\section{The Artru-Collins Asymmetry}

We  further consider  the process 
$e^+(l) e^-(l') \rightarrow (\pi^+ \pi^-)_{\rm jet1} (\pi^+ \pi^-)_{\rm jet2} X$, with (time-like) momentum transfer $q=l+l'$. Here, we have  two pairs of pions, one 
originating from a fragmenting parton and the other one from the related 
antiparton.\footnote{Variables with an extra \lq\lq bar" refer to the pair coming from the antiquark.}

The differential cross sections also depend on the 
invariant $y = P_h\cdot l / P_h \cdot q$ which is related, in the lepton 
center-of-mass frame, to the angle 
$\theta_2 = \arccos (\bm{l_{e^+}}\cdot\bm{P}_h / (|\bm{l_{e^+}}|\,|\bm{P}_h|))$, with $\bm{l_{e^+}}$ the momentum of the positron,
by $y = (1+\cos\theta_2)/2$.

The dihadron Fragmentation Functions are involved in the description of 
the fragmentation process $q\to \pi^+ \pi^- X$, where the quark has momentum $k$. 
They are extracted from the correlation function~\cite{Bacchetta:2002ux}
\begin{equation} 
\Delta^q(z,\cos\theta,M_h^2,\phi_R) = 
\frac{z |\vec R|}{16\,M_h}\int d^2 \vec k_T \; 
       d k^+\,\Delta^q(k;P_h,R) \Big|_{k^- = P_h^-/z}  \; , 
\label{eq:delta1}
\end{equation} 
where
\begin{eqnarray} 
\Delta^q(k,P_h,R)_{ij}
         & =&\sum_X \, \int
        \frac{^4\xi}{(2\pi)^{4}}\; e^{+\i k \cdot \xi}
       \langle 0|
{\cal U}^{n_+}_{(-\infty,\xi)}
\,\psi_i^q(\xi)|P_h, R; X\rangle 
\langle P_h, R;, X|
             \bar{\psi}_j^q(0)\,
{\cal U}^{n_+}_{(0,-\infty)}
|0\rangle \,.    
\label{e:delta2}
\end{eqnarray} 
Since we are going to perform the integration over the transverse momentum $\vec{k}_T$, the Wilson lines ${\cal U}$ 
can be reduced to unity using a light-cone gauge.  
The only fragmentation functions surviving after integration over the azimuthal angle defining the position of the lepton plane w.r.t. the laboratory plane~\cite{Boer:2003ya}.
\begin{eqnarray} 
D_1^q(z,\cos\theta,M_h^2) &= 4\pi\, \Tr[\Delta^q(z,\cos\theta,M_h^2,\phi_R)\,
\gamma^-],
\\
\frac{\epsilon_T^{ij}\,R_{T j}}{M_h}\, H_1^{\open\, q}(z,\cos\theta,M_h^2)
&=4\pi \, \Tr[\Delta^q(z,\cos\theta,M_h^2,\phi_R)\,i\,\sigma^{i -}\,\gamma_5].
\end{eqnarray} 
 We perform an expansion in terms of Legendre functions 
of $\cos\theta$ (and $\cos\overline{\theta}$) and  keep only the $s$- and 
$p$-wave components of the relative partial waves of the pion pair. By further 
integrating upon $d\cos\theta$ and $d\cos\overline{\theta}$, we isolate only  the  specific contributions 
of $s$ and $p$ partial waves to the respective DiFFs. 

The azimuthal Artru-Collins asymmetry $A(\cos \theta_2, z, \bar z, M_h^2, \bar M_h^2)$~\cite{Artru:1995zu} corresponds to a~$\cos(\phi_R+\phi_{\bar R})$ modulation in the cross section for the process under consideration. It can be written in terms of DiFF in the following way,
\begin{eqnarray} 
A(\cos\theta_2,z,M_h^2,\bar{z},\bar{M}_h^2) 
&= &\frac{\sin^2 \theta_2}{1+\cos^2 \theta_2} \, \frac{\pi^2}{32}\,
\frac{|\bm{R}|\,|\overline{\bm{R}}|}{M_h\,\overline{M}_h} \,
\frac{\sum_q e_q^2 \, H_{1,q}^{\open sp}(z,M_h^2)\,
          \overline{H}_{1,q}^{\open sp}(\overline{z},\overline{M}_h^2)}
     {\sum_q e_q^2\, D_{1,q}^{ss+pp}(z,M_h^2) \,
          \overline{D}_{1,q}^{ss+pp}(\overline{z},\overline{M}_h^2) } \; ,
\label{eq:asye+e-}
\end{eqnarray} 
with $ |\bm{R}| =\frac{M_h}{2} \sqrt{1- 4\,m_{\pi}^2/M_h^2} $. To extract a parameterization of the function $H_1^{\sphericalangle}$, we need to know the function $D_1$.

\section{Electron-Positron Annihilation: The Unpolarized Cross-Section from BELLE}

A model independent parameterization of a function means a huge freedom on the functional form one will choose. First, one can guess the causes of the shape of the data  from physical arguments. 
One can get inspired in comparing the model results with the data: here, we take into account the results of Ref.~\cite{Bacchetta:2006un} ---including a critical eye on its shortcomings--- in defining the shape of the MC histograms for the  unpolarized cross section.

In the process $q\to \pi^+ \pi^- X$, the prominent channels for an invariant mass of the pion pair ranging $2 m_{\pi} < M_h \lesssim 1.5$ GeV are, basically :
\begin{itemize}
\item the  fragmentation into a $\rho$ resonance  decaying into $\pi^+ \pi^-$, responsible for a peak at $M_h \sim$ 770 MeV ; 
\item the fragmentation into a $\omega$ resonance decaying into $\pi^+  \pi^-$, responsible for a small peak at $M_h \sim$ 782 MeV plus the fragmentation into a $\omega$ resonance decaying into $\pi^+  \pi^- \pi^0$ 
($\pi^0$ unobserved), responsible for a broad peak around $M_h \sim$ 500 MeV ;
\item the continuum, i.e. the fragmentation into an ``incoherent'' $\pi^+ \pi^-$ pair, is probably the most important channel. It is also the most  difficult channel to describe with purely model-based physical arguments.
\end{itemize}

In addition to the channel decomposition, one has to take into account the flavor decomposition of the cross section. This further decomposition  is  particularly important if one wants to be able to use the resulting parametrization in another context, e.g., SIDIS. 
For the time being, the MC data provided by the BELLE collaboration are additionally separated into flavors, i.e., $uds$ contributions and $c$ contributions.
The experimental analyses conclude that the charm contribution to the unpolarized cross section is non-negligible at BELLE's energy.\footnote{R.~Seidl's talk at TMD workshop, ECT$^{\ast}$, June 2010.} 

\begin{figure}
\centering
\includegraphics [height=4.5cm]{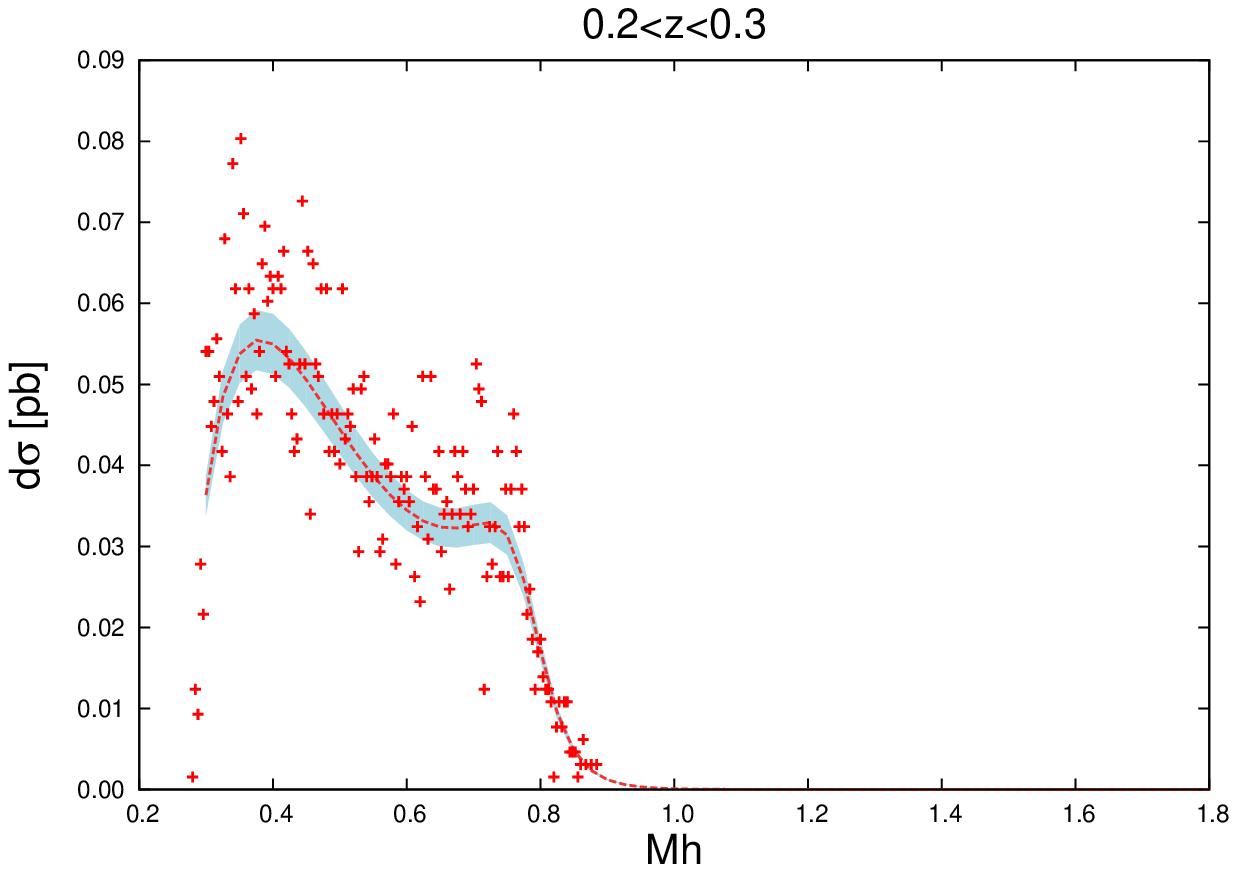}
\includegraphics [height=4.5cm]{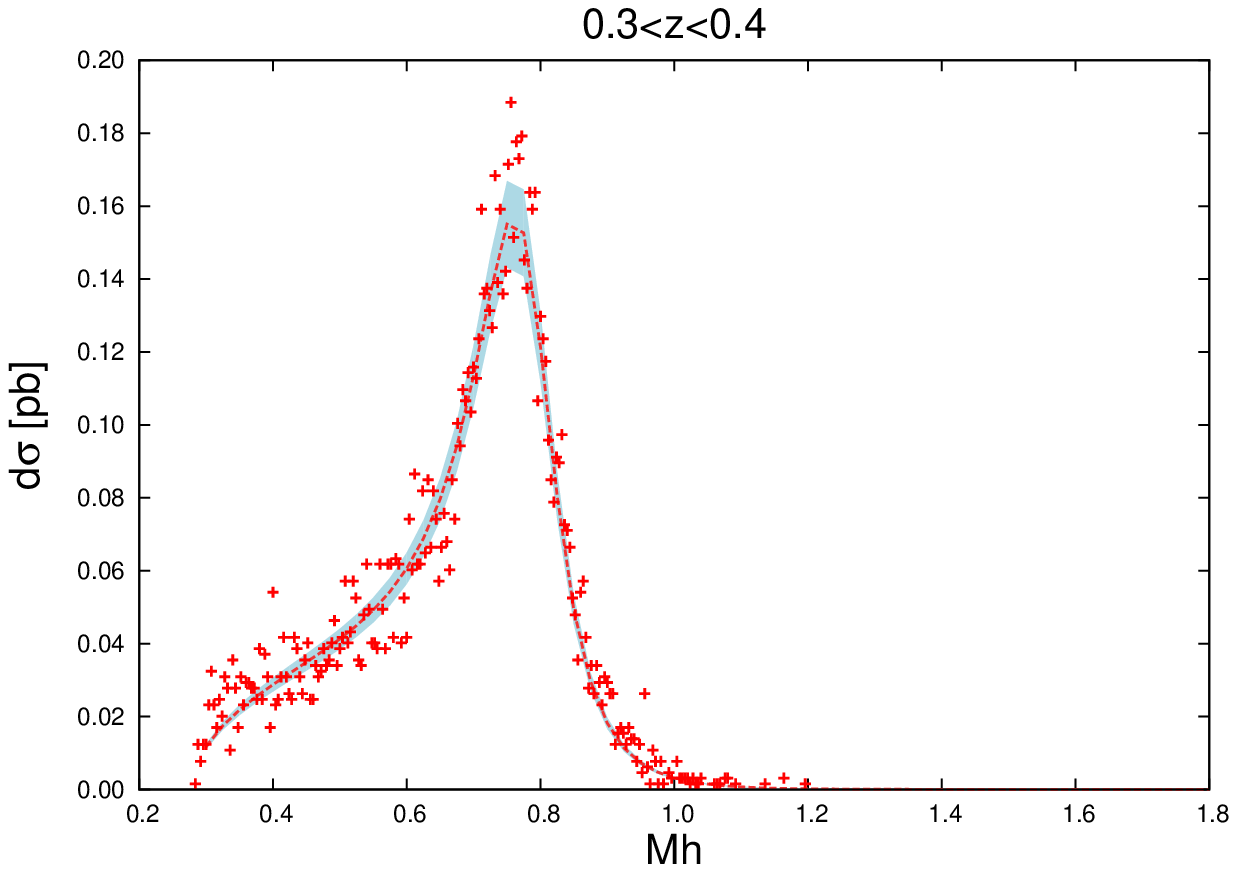}\\
 \includegraphics [height=4.5cm]{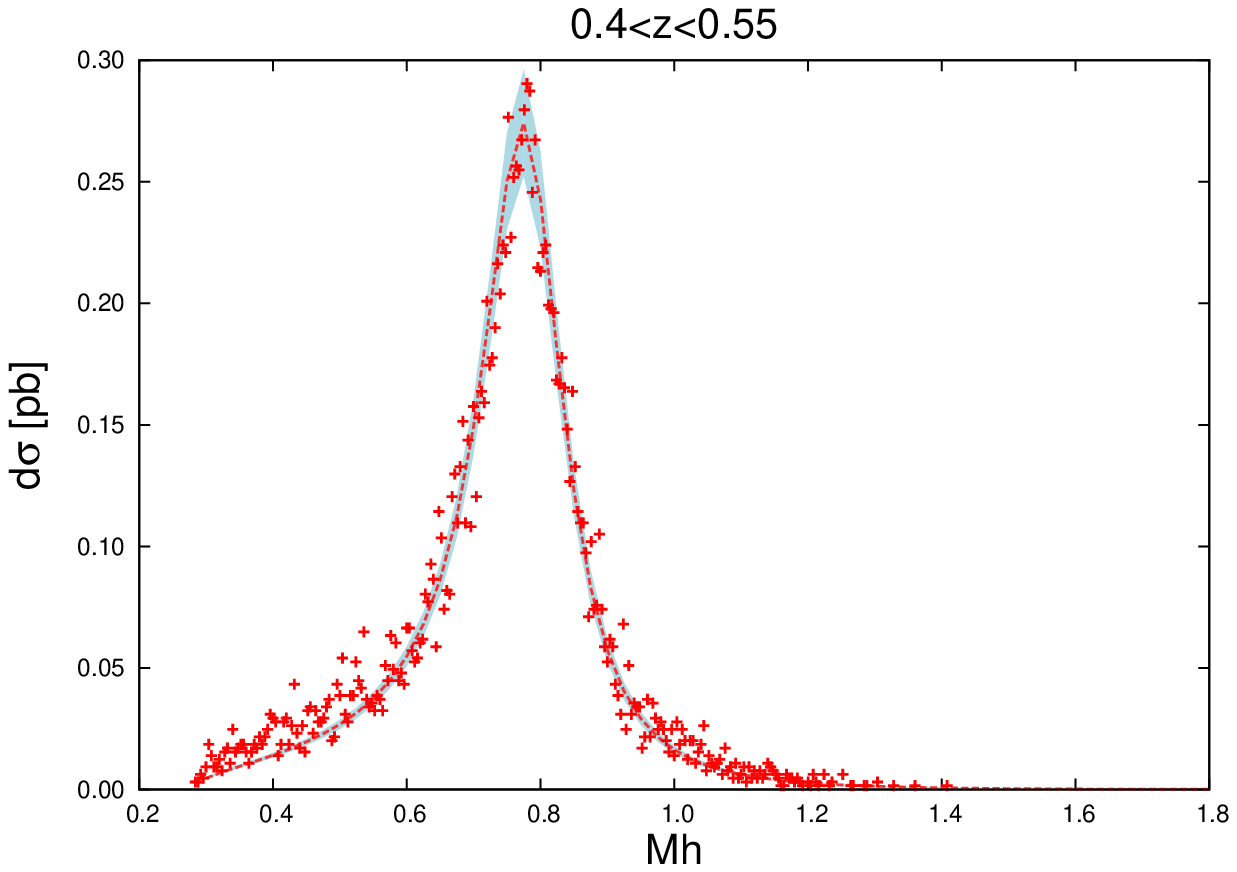} 
  \includegraphics [height=4.5cm]{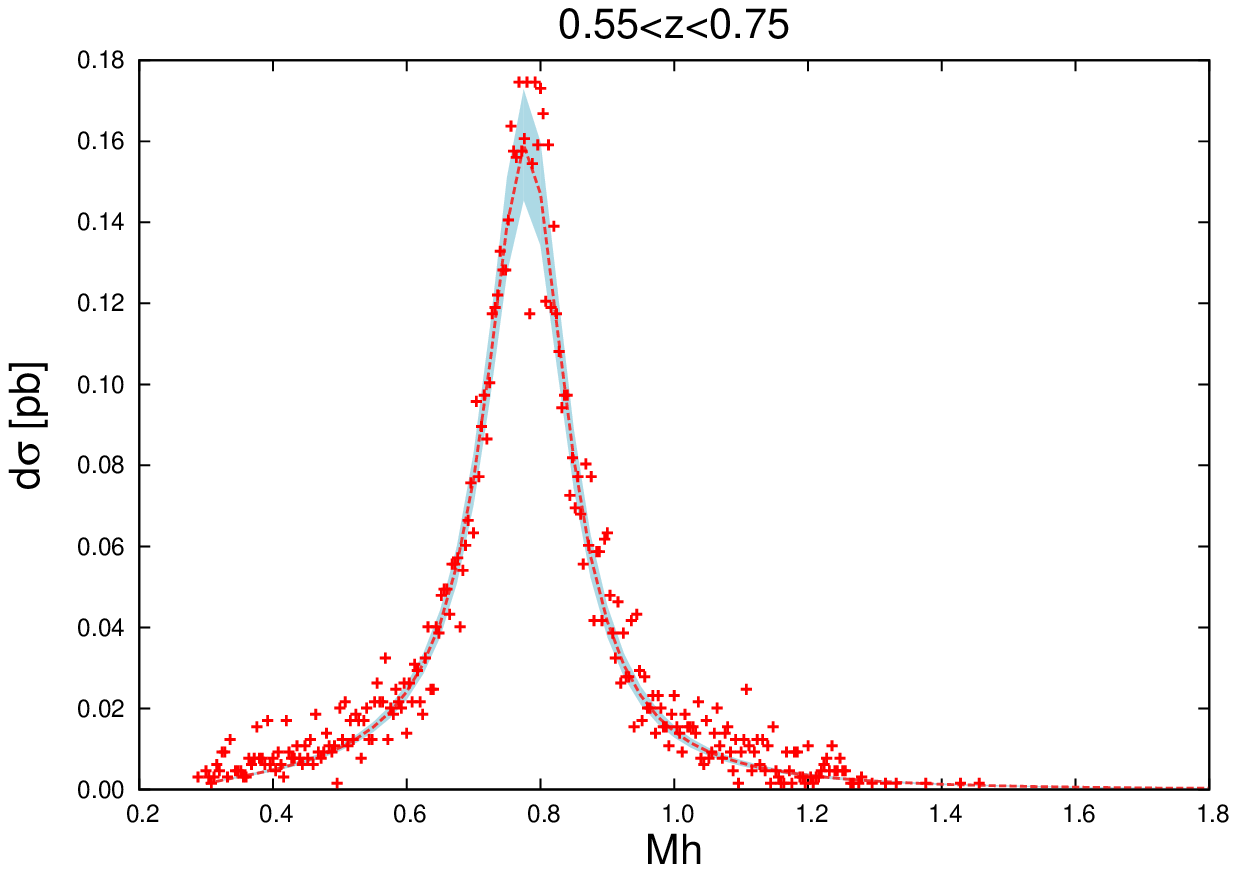} 
  \caption{ Number of events $N$ for the unpolarized $e^+e^-$ annihilation into 2 pions in  a jet (plus anything else) at BELLE, normalized by the integrated luminosity $647.26 \,\mbox{pb}^{-1}$.   We show only the resonant channel for the $\rho$ production.The data are represented by the dots. The error on the data (not plotted here) is assumed to be $\sqrt{N}$.  The dashed lines represent the parameterization, and the band its errorband. $M_h$ in GeV.}
 \label{rho_mc_d1}
\end{figure}

The main considerations one can do, before fitting the data, are the following.
First, the most important contribution from the charm is in the continuum and cannot be neglected.
The determination of a functional form for $D_1$ consists then in four parallel steps, i.e. the 2-dimensional parameterization of the $\rho$ and $\omega$ channels and of  the continuum for $uds$ and only of the continuum for the charm. 
Second,  it can be deduced that both the $\rho$ and $\omega$ channels play a role at high $z$ values, while it seems that the $\rho$ is less important at lower $z$ values, as it can be seen in Fig.~\ref{rho_mc_d1}. On the other hand, the continuum decreases with $z$, and this behavior is different for the $uds$ and the $c$ flavors. Also, it can be observed, e.g. in Fig.~\ref{rho_mc_d1}, that the behavior in $M_h$ changes from $z$-bin to $z$-bin. Those are signs that the dependence on $z$ and $M_h$ cannot be factorized.%

 Following Eq.~(\ref{eq:asye+e-}), the unpolarized cross section that we are considering here is differential in $(\cos \theta_2, z, M_h^2, \ovl z,\ovl M_h^2)$. The $\theta_2$-dependence is provided by the BELLE collaboration, and the set of variables $(\bar z, \bar M_h)$ is integrated out  within the experimental bounds.
The methodology is as follows. The unpolarized cross section, differential in $M_h$ and $z$, is
\begin{eqnarray}
 \frac{d \,\sigma^{U}}{2M_h dM_h dz}
=\sum_{a,\ovl a}\; e_a^2 \,\frac{1}{3} \frac{6\,\alpha^2}{Q^2}\,\langle 1+\cos^2\theta_2 \rangle\,\;z^2 
\,  D_1^{a} (z, M_h^2)\,\int_0^1  d{\ovl z}\,\int_{2m_{\pi}}^{M_{h}^{\mbox{\scriptsize max} }} 2  \ovl M_hd \ovl M_h\, {\ovl z}^2\, {\ovl D}_1^{a} ({\ovl z}, \ovl M_h^2)\quad ,\nn
\end{eqnarray}
with the integration limits to be modified according to the experiment, and where $D_1=D_1^{ss+pp}$.
For both the $uds$ and $c$ flavors, the fitted function takes the form
\begin{eqnarray}
FF_U&=& \frac{1}{3} \,\frac{6\alpha^2}{Q^2} \langle 1+\cos^2\theta_2 \rangle\, \sum_a\, {e_a^2}\, 
\int_{z_{bin}} d{z}
 f_{D_1}^a(z, M_h) \int_{0.2}^{1} d{\ovl z}  \int_{2m_{\pi}}^{1.5 \mbox{\tiny GeV}}  d \ovl M_h f_{D_1}^{\bar a}(\bar z, \bar M_h)\quad ,
 \label{ff}
\end{eqnarray}
where $\int_{z_{bin}} d{z}$ means that we average over the $z$-dependence each $z$-bin, and with our functional form appearing in
%
$ f_{D_1}^a(z, M_h)=2M_h\, z^2\,D_{1}^{a}(z, M_h^2)$.
%
The upper integration limit in Eq.~(\ref{ff}) is chosen to be in agreement with the condition $M_h< <Q$. The DiFFs for quarks and antiquarks are related through the charge conjugation rules described in Ref.~\cite{Bacchetta:2006un}.

The determination of a functional form $f_{D_1}$ is done by fitting, by means of a $\chi^2$ goodness-of-fit test,  the MC histograms (4 $z$-bins and about 300 $M_h$-bins) for each channel.
  The best-fit functional forms lead to interesting results. The most important point is that there is no way the $z$ and the $M_h$ dependence can be factorized.  Moreover, we have realized that no acceptable  fit would be reached  with a trivial functional form for the continua. 
In Fig.~\ref{rho_mc_d1} we show, as an example, the MC of the $\rho$ production together with its parametrization. 
In the depicted case, the joint $\chi^2/$d.o.f. is $\sim1.25$~\cite{noi-diff}. We quote the $\chi^2$ values for the other channels: $\omega$-production ($\chi^2/$d.o.f. $\sim 1.3$)~; $uds$-background ($\chi^2/$d.o.f. $\sim 1.4$)~; $c$-background ($\chi^2/$d.o.f. $\sim 1.55$)~\cite{noi-diff}.
The propagation of errors gives rise to the $1$-$\sigma$ error band shown in light blue.

\section{Towards an extraction of  $H_1^{\sphericalangle}$}

The DiFF  $H_1^{\sphericalangle}(z, M_h)$ can be extracted from the Artru-Collins asymmetry. The preliminary data from the BELLE collaboration~\cite{Vossen:2009xz} will be our starting point.  Those data are binned in $(z, \bar z)$ and $(M_h, \bar M_h)$. While we have stated in the previous section that no factorization of the $(z, M_h)$ variables is possible for $D_1$, the data do not allow us to make a similar statement for $H_1^{\sphericalangle}$.

The next step consists in the  determination of a  functional form, e.g.,
\begin{eqnarray}
 f_{H_1^{\sphericalangle}}(z, M_h,\bar  z, \bar M_h)&\propto& f(z)f(\bar z)\, g(M_h)g(\bar M_h)\quad .
 \label{simple}
\end{eqnarray}
Even if we expect the $H_1^{\sphericalangle}$ to arise from an $sp$-wave interference, we presently have no guidance on the interplay of the $(z, M_h)$ variables in the asymmetry.
We opt for the  simpler  functional form~(\ref{simple}) instead. 
 Given the large uncertainties ---we sum statistical and systematic errors in quadrature--- on the asymmetry as well as the shape of the $(z, \bar z)$ dependence, it is easily realized that more than one functional form could  fit the data. We are currently working in improving our fitting procedure in order to get as much information as we can from the data.

Once we will have  determined the $z$ as well as the $M_h$-dependence of the $H_1^{\sphericalangle}$ DiFF, we will have to face the flavor decomposition problem. This step will crucially influence the extraction of transversity, see Eq.~(\ref{eq:asydis}).

We conclude by highlighting the importance of DiFFs in the extraction of transversity. We are eagerly looking forward to analyzing the published data on $e^+e^-$ from the BELLE collaboration and to going through the described methodology.


\ack

We are thankful to the BELLE collaboration for useful information on the data. 

\end{document}